\documentclass[useAMS,usenatbib]{mn2e}

\usepackage{color}
\usepackage{amssymb}
\usepackage{epsfig}

\newcommand{\msun}{\mbox{$M_\odot$}}    
\newcommand{\OH}{\mbox{$\rm[O/H]$}}    
\newcommand{\FeH}{\mbox{$\rm[Fe/H]$}}
\newcommand{\OFe}{\mbox{$\rm[O/Fe]$}}
\newcommand{\Gyr}{\,{\rm Gyr}}
\newcommand{\Myr}{\,{\rm Myr}}
\newcommand{\kpc}{\,{\rm kpc}}
\newcommand{\dex}{\,{\rm dex}}
\newcommand{\MH}{\mbox{$M_{\rm H}$}}
\newcommand{\MO}{\mbox{$M_{\rm O}$}}
\newcommand{\MFe}{\mbox{$M_{\rm Fe}$}}

\newcommand{\AB}{$\rm A_\bullet$}
\newcommand{\BB}{$\rm B_\bullet$}
\newcommand{\CB}{$\rm C_\bullet$}
\newcommand{\DB}{$\rm D_\bullet$}
\newcommand{\EB}{$\rm E_\bullet$}
\newcommand{\FB}{$\rm F_\bullet$}
\newcommand{\GB}{$\rm G_\bullet$}
\newcommand{\HB}{$\rm H_\bullet$}
\newcommand{\IB}{$\rm I_\bullet$}
\newcommand{\JB}{$\rm J_\bullet$}
\newcommand{\DMM}{$\rm D^{--}$}
\newcommand{\DM} {$\rm D^-$}
\newcommand{\DP} {$\rm D^+$}
\newcommand{\DPP}{$\rm D^{++}$}

\newcommand{\cm}{\checkmark}

\begin{document}
\title[Chemical enrichment in isolated barred spiral galaxies.]
{Chemical enrichment in isolated barred spiral galaxies.}

\author[Hugo Martel et al.]
{Hugo Martel,$^{1,2}$
Christian Carles,$^{1,2}$
Fid\`ele Robichaud,$^{1,2}$
Sara L. Ellison,$^3$\newauthor
and David J. Williamson,$^4$\\
$^1$D\'epartement de physique, de g\'enie physique et d'optique,
Universit\'e Laval, Qu\'ebec, QC, G1V 0A6, Canada\\
$^2$Centre de Recherche en Astrophysique du Qu\'ebec, C. P. 6128,
Succ. Centre-Ville, Montr\'ea, QC H3C 3J7Q, Canada\\
$^3$Department of Physics and Astronomy, University of
Victoria, Victoria, BC V8P 1A1, Canada\\
$^4$Department of Physics and Astronomy,
University of Southampton, Southampton SO17 1BJ, UK}

\date{Accepted XXX. Received XXX; in original form XXX}

\pagerange{\pageref{firstpage}--\pageref{lastpage}} \pubyear{XXX}

\maketitle

\label{firstpage}

\begin{abstract}
To investigate the role of bars in the chemical evolution of
isolated disc galaxies,
we performed a series of 39 gas dynamical simulations of isolated barred
and unbarred galaxies with various masses, initial gas fractions, and AGN
feedback models. The presence of a bar drives a substantial amount of gas
toward the central region of the galaxy. In the most massive galaxies,
this results in a violent starburst, followed by a drop in star formation
resulting from gas exhaustion. The time delay between
Type~Ia and Type~II supernovae explosions means that barred galaxies experience
a rapid increase in \OH\ in the central region, and a much more gradual
increase in \FeH.
In unbarred galaxies, star formation proceeds
at a slow and steady rate, and oxygen and iron are produced at steady rates
which are similar
except for a time offset.
Comparing the abundance ratios in barred and unbarred galaxies
with the same central stellar mass $M_*$, we find in barred galaxies
an enhancement of $0.07\dex$ in \OH, $0.05\dex$ in \FeH, and
$0.05\dex$ in $\OFe$. The \OH\ enhancement is in
excellent agreement with observations from the {\sl SDSS}.
The initial gas fraction has very
little effect on
the abundance ratios in barred and unbarred galaxies, unless the
galaxies experience a starburst.
We considered
AGN-host galaxies located near the bottom of the AGN regime,
$M_*\gtrsim3\times10^{10}\msun$,
where AGN feedback dominates over supernovae feedback.
We found that the impact of
AGN feedback on the central abundances is marginal.
\end{abstract}


\section{INTRODUCTION}

In the Cold Dark Matter cosmological scenario of structure formation,
present-day galaxies
result from the merger of smaller galaxies, and from accetion of
intergalactic matter. These processes can greatly affect the chemical
properties of the resulting galaxies. There is evidence, both observationally
\citep{kgb06,kewleyetal10,ellisonetal08a,rkc10,richetal12,sanchezetal14}
and through numerical simulations
\citep{dimatteoetal09,
montuorietal10,rkb10,perezetal11,silleroetal17} that major mergers
(progenitors with mass ratio $\sim\hbox{1:1}$) cause low-metallicity gas
to flow toward the centre of the galaxies. This lowers the central
abundances of metals, flattens the radial metallicity gradient, and could
potentially explain why some massive galaxies
fall below the observed mass-metallicity relation
\citep{ellisonetal08a,kewleyetal10,scudderetal12}.

While some galaxies might experience major mergers and significant accretion
up to the present epoch, others, in particular galaxies located in
low-density environments, might complete most of their mass assembly at
higher redshift. If, at that point, the mass gained by mergers and accretion
becomes negligible, and tidal perturbations induced by encounters with other
galaxies also become negligible, the galaxies can be considered as isolated.
\citet{hopkinsetal10} have used an analytical approach to determine under
what conditions galaxies can be considered as isolated.
Using multi-zoom cosmological simulations,
\citet{lcs12} showed that galaxies can experience a wide range of
mass assembly histories, with some galaxies growing all the way to
the present, and others completing their mass assembly before redshift
$z=1$.

If a galaxy completes its mass assembly before the present, its
subsequent evolution will be essentially secular. The physical processes
driving secular evolution differ from the ones involved during mergers.
While, in both cases, the flow of matter toward the centre of galaxies
drives the chemical evolution, the origin of this flow is different.
Simulations of merging galaxies show that the mass ratio and orbital
geometry are the main factors determining the evolution and the resulting
chemical abundances and distributions. In isolated galaxies, secular
evolution will depend on intrinsic properties of the galaxy, both
structural and kinematic, the most important one being the presence of
absence of a bar.
The torque exerted by a bar transports
angular momentum from the inner to the outer regions of the
galaxy \citep{ds00,ks12,lokasetal14,seideletal15}, leading to a
redistribution of the gaseous and stellar component \citep{gda01,gkc15}.
Gas flows from the outer to the inner regions of the galaxy and
accumulates in the central region
\citep{cg85,ce93,mtss02,rt04,bsw10,mastersetal12,kpa15}.
This accumulation of gas can cause an increase in star formation
\citep{devereux87,martin95,mf97,ak01,huntetal08,cg11},
chemical enrichment \citep{fbk94,fb95}, and AGN activity
\citep{sfb89,sn93,hs94,combes03,jogee06}.

To assess observationally the effect of bars on
the secular evolution of isolated galaxies,
we must choose an observable physical property
that is predicted to be affected by the presence
of bars. This could be the SFR, chemical abundances in
the gas phase, chemical abundances in the stellar phase,
abundance gradients, or AGN luminosity. We then compare barred
and unbarred galaxies to look for variations
that would result from the presence of the bar.
However, these various quantities can vary greatly
both among barred and unbarred galaxies. This leads to the issue of
pairing, that is, determining which barred and unbarred galaxies should be
compared to one another.
To make meaningful one-to-one comparisons,
one must select galaxies that are physically
similar, except for the
presence of the bar. Ideally, one would like to compare galaxies with the
same total mass, or equivalently the same rotational velocity.
However, this information is often not available.
Instead, one must use a proxy such as the stellar mass, baryonic mass,
or magnitude.

While several observational studies
indeed show increases in SFR, metallicity,
and AGN activity in barred galaxies, others have yielded
conflicting results, suggesting that the details of the various physical
processes involved are not fully understood.
Several observational studies of the SFR in barred galaxies
found no increase \citep{pr90,mf97,ccd99,willettetal15}
or an increase only in early-type spiral galaxies \citep{hfs97,jbk09}.
\citet{cheungetal13}
showed that the specific star formation rate (SSFR) is
anti-correlated with the presence of a bar.

Observations of central metallicities in barred galaxies
have also yielded conflicting results.
For the gaseous component,
\citet{hw99} and \citet{dr99}
compared barred and unbarred galaxies with the same magnitude
$M_B$, and found that in general barred galaxies have flatter metallicity
gradients and lower central metallicities than unbarred ones.
\citet{considereetal00} did a similar
study, and found that bars have little impact on the
star formation history and chemical evolution of starburst
galaxies, and suggest that the bars in their sample are too young.
More recently, \citet{kaplanetal16} selected eight
spiral galaxies from the VENGA survey, and found that isolated
barred and unbarred galaxies exhibit similar metallicity profiles
up to large radii.
\citet{ellisonetal11} studied the SFR and gas metallicity
in {\it the central region\/} of
barred and unbarred galaxies,
as captured by the 3 arcsecond {\sl SDSS\/} fibre.
The {\sl SDSS\/} aperture size corresponds to a radius of
approximately $1\kpc$ at $z\sim0.03$.
By comparing
barred and unbarred galaxies with the same central stellar mass
$M_*$, they found that
the SFRs of barred galaxies are enhanced by $0.2\dex$ compared to
unbarred galaxies, but only for galaxies with
$\log(M_*/\msun)>10$.
In contrast, the central gas metallicity is enhanced by $0.05\dex$
in barred galaxies, at all stellar masses.
However,
\citet{cachoetal14} also studied barred galaxies in the {\sl SDSS\/}
and found
no significant differences in metallicity.
These authors argue that differences in sample selection and metallicity
indicator explain the discrepancy between there results and those of 
\citet{ellisonetal11}.

The metallicity of the stellar component has received
less attention. \citet{mh06} and \citet{psb11} compared barred and unbarred
galaxies with the same rotational velocity or velocity dispersion, and found
an enhancement in stellar metallicity in barred galaxies, though the
differences might be too small to
be statistically significant.
The study of \citet{cg11,cachoetal14}, and \citet{sanchez-blazquezetal14}
found no significant differences between barred
and unbarred galaxies.

These observational results illustrate the complexity of the chemical
evolution of galaxies. Numerous factors must be accounted for, including
galaxy mass, morphological type, gas mass fraction, bar-strength and size,
and the role of AGN. Numerical simulations can help
shed a light on the physical processes driving chemical evolution
and how they are affected by the presence of a bar.
In a previous paper (\citealt{paper1}, hereafter Paper~I)
we presented a numerical study of star formation and metallicity
enrichment in barred and unbarred galaxies. We showed that there is
no direct connection between central star formation and central
metallicity, because an important fraction of the central metals were
produced by stars that formed outside of the central region.
In two follow-up papers (\citealt{paper2}, hereafter Paper~II;
\citealt{paper3}, hereafter Paper~III), we presented
more extensive numerical studies
of the star formation history in barred and unbarred galaxies,
focussing on the effect of the total mass and initial gas fraction
(Paper~II) and the effect of feedback resulting from
the presence of a central AGN (Paper~III).
In this paper, we are revisiting the simulations
presented in Papers~II and~III, this time focusing
on the history of chemical enrichment. This provides
us with a much more extended set of simulations than the one previously
used in Paper~I (39 vs. 4).
Our goal is to investigate
the difference in metal production and distribution 
in barred and unbarred galaxies, the dependence on stellar mass, and the
role played by AGN feedback, to ultimately understand the interplay between
the various physical processes responsible for chemical enrichment
in disc galaxies.

The remainder of this paper is organised as follows: In section~2, we
describe our numerical algorithm. In section~3, we present our suite of
simulations. Results are presented in section~4.
Summary and conclusions are presented in section~5.

\section{THE NUMERICAL ALGORITHM}

\subsection{The GCD+ algorithm}

The simulations were performed using the numerical algorithm
GCD+ \citep{kg03,rk12,bkw12,kawataetal13,kawataetal14,wmk16}.
GCD+ is a particle-based, gas-dynamical algorithm
specially designed to simulate galactic chemodynamical evolution.
The algorithm tracks the dynamics and chemical evolution of
the stellar and gaseous components, taking into account
star formation, supernova feedback, metal enrichment, and diffusion.
The algorithm was recently modified to include a subgrid treatment
of AGN accretion and feedback (Paper~III).

In all simulations, the gravitational softening length is fixed at
$90\,\rm pc$. Smoothing length are adjusted such that each gas particles
has $\approx58$ neighbours. However, we impose a minimum smoothing
length of $90\,\rm pc$, to be consistent with the resolution of the
gravitational field. In simulations with a central AGN,
the smoothing length of the particle representing the central black hole
is adjusted such that the black hole has $\approx70$ particles within its
zone of influence.

\subsection{Supernova feedback and chemical enrichment}

In GCD+, the gaseous and stellar components are represented
by gas particles and star particles, respectively.
Each star particle represents an entire population of stars of a
particular mass range.
Different star particles represent stars of different mass
ranges, and together they reproduce
a \citet{salpeter55} initial mass function
(see \citealt{kawataetal14} for details).
When stars reach the end of their life, they
deposit energy and metal-enriched gas into the surrounding gas.
The algorithm calculates the amount of energy, and the mass and composition
of the ejecta. That energy and ejecta is then deposited onto the
gas particles located within one SPH smoothing length of the star particle.
We calibrate the supernova feedback using the parameters given in
\cite{rk12}: each SN produces an energy $E_{\rm SN}=10^{51}\rm erg$,
with only 10\% of that energy contributing to feedback; the remainder
is lost through radiation.

The algorithm include chemical enrichment from Type~II
and Type~Ia SNe, using the yields tabulated from \citet{ww95}
and \citet{iwamoto99}, respectively.
We include the elements H, He, C, N, O, Ne, Mg,
Si, and Fe. Hydrogen and helium are primordial elements that are produced
primarily during the big bang. Carbon, nitrogen, oxygen, neon,
magnesium, and silicon are produced by
the explosion of massive stars in the form of Type~II SNe.
The lifetime of these stars being very short
(5 to $10\Myr$), the delay between star formation and $\alpha$-element
enrichment of the ISM is also very short \citep{tsujimotoetal95}.
Iron is produced primarily by Type~Ia SNe, whose progenitors have lifetimes
of order $1\Gyr$, and therefore there is an important delay between the
time when star formation takes place and the time when the ISM is enriched
in iron.

We note that star particles inherit the metallicity of the interstellar
gas out of which they form. Hence, the code keeps track of the metallicity
both in the stellar and in the gaseous components. In this paper, we focus
on the metallicity of the gaseous component, which we will simply
refer to as ``the metallicity.''

\subsection{AGN feedback and dynamics}

Feedback from a central supermassive black hole is a recent addition to GCD+,
which was described in Paper~III. We refer the reader to \citet{wt13}
and Paper~III for details. In this implementation, the accretion rate
onto the supermassive black hole is calculated analytically, using the
Bondi accretion rate \citep{hl39,bh44,bondi52}, with an upper limit
imposed by the condition that the accretion rate should
never exceed the Eddington luminosity.

With this accretion rate, the mass of the black hole increases continuously
during the course of the simulation. However, the GCD+ algorithm imposes a
numerical constraint on the black hole mass: it can only be a multiple
of the mass of the gas and star particles. To handle this situation,
the mass of the black hole is represented by two values, a dynamical mass
$M_\mathrm{dyn}$, and an internal sub-grid mass $M_\mathrm{SGS}$.
The sub-grid mass is calculated using the analytical accretion rate,
and its value is used to calculate the accretion rate itself, and the
strength of the AGN feedback. The dynamical mass is used for the
calculation of the gravitational field.
In order to maintain a quasi-equality between the two masses, the
dynamical mass increases discontinuously with time, by accreting gas
particles which are then removed from the calculation. The mass accreted is
used to calculate the amount of feedback energy that will be deposited into
the surrounding ISM. That
energy is divided evenly amongst all gas particles in the vicinity of
the black hole, using a combination of thermal and kinetic feedback
(see, e.g., \citealt{baraietal14}).

\section{THE SIMULATIONS}

The simulations were described in Papers II and III.
In Paper~II, we presented simulations of barred and unbarred galaxies of 
various masses and initial gas fractions,
without AGN feedback. In Paper~III, we presented simulations
of massive barred and unbarred galaxies, with and without an AGN.
These papers focused on the star formation history of galaxies.
In this paper, we revisit these simulations, this time focusing on the
metallicity evolution. This section presents a brief description of the
simulations. For a more detailed description, we refer the reader to
Papers~II and~III.

\subsection{Initial conditions}

Initial conditions are generated using the technique described in
\citet{gkc15}. For each simulation, we first specify an initial stellar
mass $M_*$. The initial gas mass and dark matter mass are then determined
using the scaling relations of \citet{belletal03}, \citet{coxetal06},
and \citet{mosteretal10}, respectively. The proportion of stellar and
gaseous mass is then adjusted for some of the runs which are
either gas-poor or gas-rich (see Table~1 below). We assume a
static dark matter halo described by a NFW profile with a concentration
parameter $c$. For barred galaxies, we use $c=8$. For unbarred galaxies,
we use $c=20$, which has the effect of suppressing the bar instability.

The stellar and gaseous components are represented with equal-mass particles.
In the initial conditions, the stellar and gas
particles form axisymmetric discs
with exponential surface density profiles. The scale length and scale height
of the stellar disc are fixed by observational constraints
(see, e.g., \citealt{shenetal03}).
The scale length of the gas disc is twice as large as the one
of the stellar disc. The scale height of the gaseous disc
is determined by the condition of hydrostatic equilibrium.
Particles are given an initial velocity equal to the circular velocity at
their location.

Both discs are given an initial metallicity gradient, using
\begin{eqnarray}
\FeH &=0.2-0.05R\,,\\
{\rm[\alpha/Fe]}&=-0.16\FeH(R)\,.
\end{eqnarray}

\noindent
where $R$ is in kpc. $\alpha-$elements are initially present in the
stellar component only.
We modify the metallicity of each particle by adding a gaussian scatter of 
$0.02\,\rm dex$ to create a local dispersion of their abundances. The star
particles are assigned an initial age using an age-metallicity relation 
$\FeH=-0.04\times\,\rm age(Gyr)$. This
approach is somehow ad hoc since isolated
galaxies of different masses might have had different abundance profiles when
they complete their mass assembly and enter their secular evolution regime,
not to mention possible variations amongst galaxies of a given mass.
For these reasons, we focus on the changes of abundances during the
secular evolution regime.
A similar argument was made in \citet{silleroetal17}.

Finally, for simulations with AGN feedback, the particle representing the
central black hole is located at the centre of mass of the galaxy.
The black hole masses (dynamic and subgrid)
are initialised at $M_{\rm dyn,i}=M_{\rm SGS,i}=10^6\msun$.

\subsection{Runs and parameters}

The parameters of the simulations are listed in Table~\ref{initial}.
Galaxies O--I are taken from Paper~II.
The second, third, and fourth column give the
initial stellar mass, initial gas mass, and dark matter halo mass, 
respectively. The fifth and sixth columns indicate whether a bar will
form by instability, or bar formation will be prevented by the
use of a dark matter halo with a high concentration parameter.
For most combinations of physical parameters,
we simulated a barred and an unbarred galaxy.
For instance, we will refer to the two galaxies A
as ``galaxy A barred'' and ``galaxy A unbarred.''
The seventh, eighth, and ninth
columns give the initial number of star particles,
gas particles, and the initial
gas fraction, respectively.
Note that the high-mass galaxy I unbarred was included in the set of
simulations in order to have an unbarred galaxy that has, at late time,
a central stellar mass similar to that of galaxy H barred.

Galaxies $\rm D^{--}$, $\rm D^-$, $\rm D^+$, and $\rm D^{++}$
are also taken from Paper~II, and have the same
initial stellar mass as galaxies~D, but different initial gas masses,
and therefore different initial gas fractions.

\begin{table*}
\centering
\begin{minipage}{140mm}
\caption{Initial properties of the simulated galaxies.}
\begin{tabular}{@{}lrrrccrrrrrl@{}}
\hline
Galaxy & $M_*^a$ & $M_{\rm gas}^a$ & $M_{200}^a$ &
barred & unbarred & $N_*$ & $N_{\rm gas}$ & $f_{\rm gas}$ & 
$t_{\rm AGN}^b$ & $f_{\rm kin}$ & Colour \\
\hline
O     &  4.00 &  1.72 &  265 & \cm & \cm &  52 501 &  22 546 & 0.300
      & $\cdots$ & $\cdots$ & Purple \\
A     &  5.00 &  2.04 &  299 & \cm & \cm &  66 584 &  27 244 & 0.289
      & $\cdots$ & $\cdots$ & Magenta \\
B     &  6.30 &  2.45 &  341 & \cm & \cm &  85 121 &  33 078 & 0.279
      & $\cdots$ & $\cdots$ & Blue \\
C     &  7.90 &  2.92 &  389 & \cm & \cm & 108 210 &  40 008 & 0.269
      & $\cdots$ & $\cdots$ & Cyan \\
D     & 10.00 &  3.51 &  450 & \cm & \cm & 138 869 &  48 749 & 0.259
      & $\cdots$ & $\cdots$ & Green \\
E     & 12.50 &  4.18 &  519 & \cm & \cm & 175 774 &  58 748 & 0.250
      & $\cdots$ & $\cdots$ & Dark green \\
F     & 15.80 &  5.02 &  609 & \cm & \cm & 225 010 &  71 426 & 0.241
      & $\cdots$ & $\cdots$ & Lime \\
G     & 20.00 &  6.03 &  726 & \cm & \cm & 288 333 &  86 901 & 0.231
      & $\cdots$ & $\cdots$ & Orange \\
H     & 25.00 &  7.17 &  872 & \cm & \cm & 364 460 & 104 583 & 0.222
      & $\cdots$ & $\cdots$ & Red \\
I     & 50.00 & 12.30 & 1848 & $\times$ & \cm & 752 656 & 185 430 & 0.197
      & $\cdots$ & $\cdots$ & Black \\
\hline
\DMM\ & 10.00 &  2.50 &  450 & \cm & \cm & 138 869 &  34 717 & 0.200
         & $\cdots$ & $\cdots$ & Purple \\
\DM\  & 10.00 &  2.98 &  450 & \cm & \cm & 138 869 &  41 480 & 0.229
         & $\cdots$ & $\cdots$ & Blue \\
\DP\  & 10.00 &   4.10 & 450 & \cm & \cm & 138 869 &  50 721 & 0.290
         & $\cdots$ & $\cdots$ & Lime \\
\DPP\ & 10.00 &  4.70 &  450 & \cm & \cm & 138 869 &  65 350 & 0.319
      & $\cdots$ & $\cdots$ & Red \\
\hline
\AB\  & 58.00 & 13.80 & 2 306 & \cm & \cm     & 514 541 & 122 694 & 0.192
      & $\cdots$ & $\cdots$ & Black/Red$^c$ \\
\BB\  & 58.00 & 13.80 & 2 306 & \cm & $\times$ & 514 541 & 122 694 & 0.192
      & 0        & 0        & Blue       \\
\CB\  & 58.00 & 13.80 & 2 306 & \cm & $\times$ & 514 541 & 122 694 & 0.192
      & 0        & 0.1      & Green     \\
\DB\  & 58.00 & 13.80 & 2 306 & \cm & \cm     & 514 541 & 122 694 & 0.192
      & 0        & 0.2      & Red     \\
\EB\  & 58.00 & 13.80 & 2 306 & \cm & $\times$ & 514 541 & 122 694 & 0.192
      & 0.5      & 0.1      & Cyan      \\
\FB\  & 58.00 & 13.80 & 2 306 & \cm & $\times$ & 514 541 & 122 694 & 0.192
      & 0.5      & 0.2      & Lime     \\
\GB\  & 64.90 &  6.90 & 2 306 & \cm & $\times$ & 575 888 &  61 347 & 0.096
      & $\cdots$ & $\cdots$ & Dark green\\
\HB\  & 64.90 &  6.90 & 2 306 & \cm & $\times$ & 575 888 &  61 347 & 0.096
      & 0        & 0.2      & Dark green     \\
\IB\  & 68.35 &  3.45 & 2 306 & \cm & $\times$ & 606 561 &  30 674 & 0.048
      & $\cdots$ & $\cdots$ & Purple \\
\JB\  & 68.35 &  3.45 & 2 306 & \cm & $\times$ & 606 561 &  30 674 & 0.048
      & 0.5      & 0.2      & Purple     \\
\hline
\end{tabular}
\label{initial}
\end{minipage}
\medskip
\begin{minipage}{140mm}
$^a$ All masses in units of $10^9\msun$.\\
$^b$ Time in Gyr when AGN accretion and feedback is turned on.\\
$^c$ The colour red is used for galaxy \AB\ barred in some
figures of section 4.5, as indicated in the captions.
\end{minipage}
\end{table*}

Galaxies \AB--\JB\ are taken from Paper~III. 
We added the bullet to their name to identify them. These galaxies all
contain a central AGN except galaxies \AB\ barred, \AB\
unbarred, \GB, and \IB.\footnote{In these simulations, we include the
black hole particle of mass $M_{\rm BH}=10^6\msun$, but turn off
accretion and feedback.}
The tenth column of Table~\ref{initial} gives the time when AGN
accretion and feedback is turned on, and the eleventh column gives the
fraction of energy feedback that goes into kinetic energy, the remainder
going into thermal energy.
These galaxies have the same dark matter halo mass
$M_{200}$ and the same total baryonic mass $M_*+M_{\rm gas}$.
Galaxies \AB--\FB\ have an initial gas fraction consistent
with the relations of \citet{belletal03} and \citet{coxetal06},
while galaxies \GB\ and \HB\ are ``gas-poor'' and
galaxies \IB\ and \JB\ are ``very-gas-poor.''
Note that in Paper~III galaxies \AB\ unbarred and
\DB\ unbarred were named K and L, respectively.

Finally, the last column of Table~\ref{initial} indicates the
colours that are used in the figures to identify the various runs.

\section{RESULTS}

\subsection{Central properties}

\subsubsection{Star formation rate}

\begin{figure}
\includegraphics[scale=0.45]{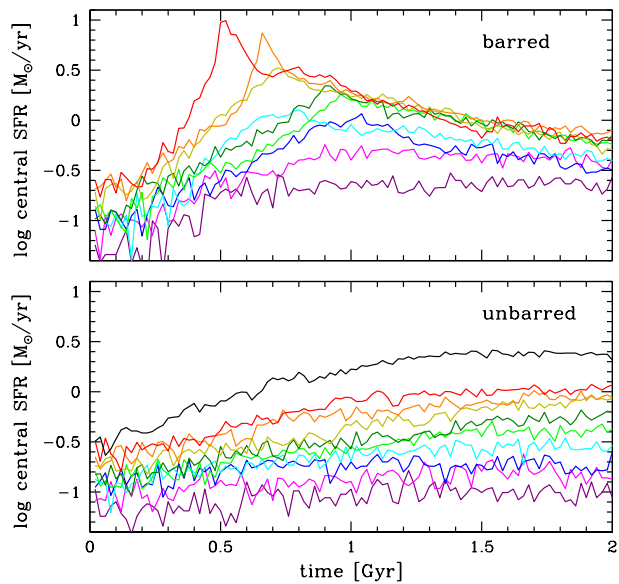}
\caption{Evolution of the central SFR in barred galaxies O--H (top) and
unbarred galaxies O--I (bottom)
The colours identify the various runs
(see last column of Table~\ref{initial}).}
\label{SFRF}
\end{figure}

Figure~\ref{SFRF} shows the central SFR in galaxies O -- I
as a function of time,
where the central region is defined as a region of radius $1\kpc$ (these
results were presented in greater detail in Paper II, but we review the SFR
properties here for context). All of the galaxies start with a low central
SFR. In barred galaxies, the central SFR rapidly increases as gas flows
inwards along the bar, peaking at $0.5-1$ Gyr before dropping away or
plateauing at later times. For more massive galaxies, the peak is earlier
and more dramatic. Such a starburst is not present in the unbarred
galaxies, and the central SFR steadily increases as gas slowly flows
toward the central region. 

\subsubsection{Metal abundances in the gas phase}

\begin{figure*}
\includegraphics[scale=0.8]{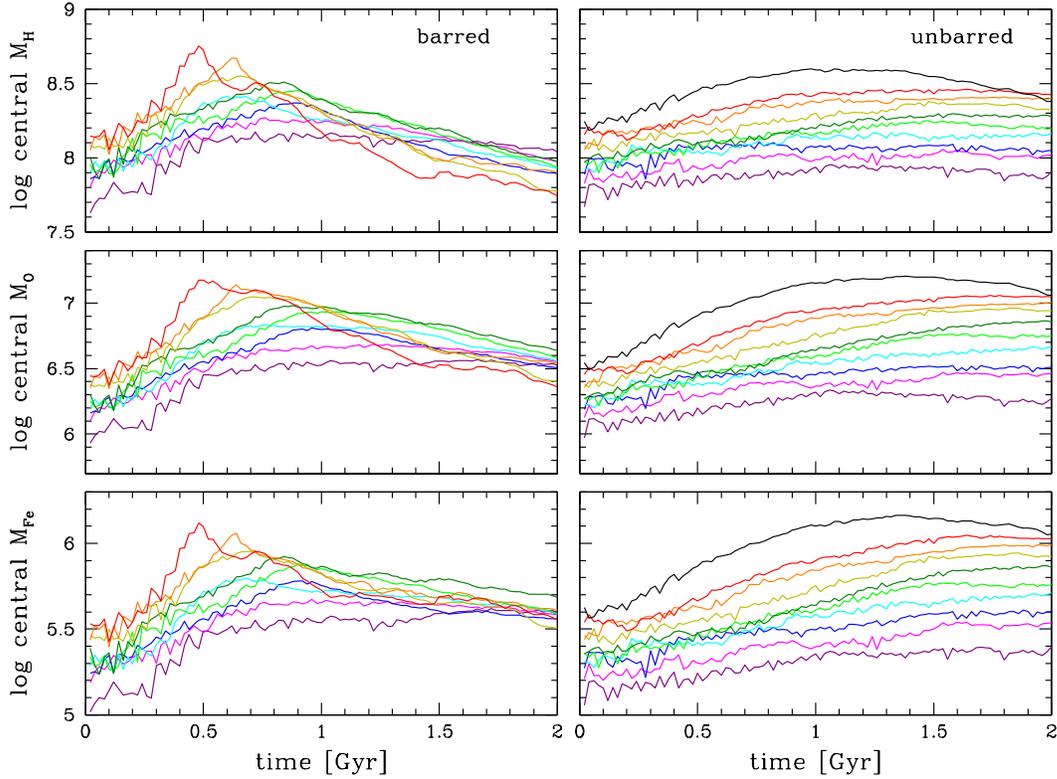}
\caption{Evolution of the central
hydrogen mass $M_{\rm H}$ (top), oxygen mass $M_{\rm O}$ (middle), and
iron mass $M_{\rm Fe}$ (bottom)
in the gas phase. Left panels: barred galaxies O--H;
right panels: unbarred galaxies O--I. The colours identify the various runs
(see last column of Table~\ref{initial}).}
\label{central_masses}
\end{figure*}

Fig.~\ref{central_masses} shows the evolution of the central hydrogen mass
\MH, oxygen mass \MO, and iron mass \MFe\ in galaxies. This is an extension of
Fig. 6. in Paper~II, which showed only the central hydrogen mass for the same
simulations. Here we find that the central oxygen and iron masses follow a
similar evolution to the central hydrogen mass. In the barred galaxies, all
the masses of all three elements increase as gas flows inwards, reaching a
peak at $0.5-0.7$ Gyr before decreasing through consumption by star
formation. This depletion slows as gas is consumed and star formation is
reduced. The more massive galaxies reach a higher peak masses, but are
depleted more quickly due to their rapid star formation. At late times,
the most massive galaxies have a smaller central metal mass than some of
the less massive galaxies.

Again, unbarred galaxies show a much smoother evolution.
The mass of all three elements steadily increases, and the values increase
with the mass of the galaxy. The only exception is the very-massive galaxy I
(black lines). The very high SFR in this galaxy results in gas starvation
at late times, causing a drop in the masses of all three elements.

\subsubsection{Metallicity}

\begin{figure*}
\includegraphics[scale=1.0]{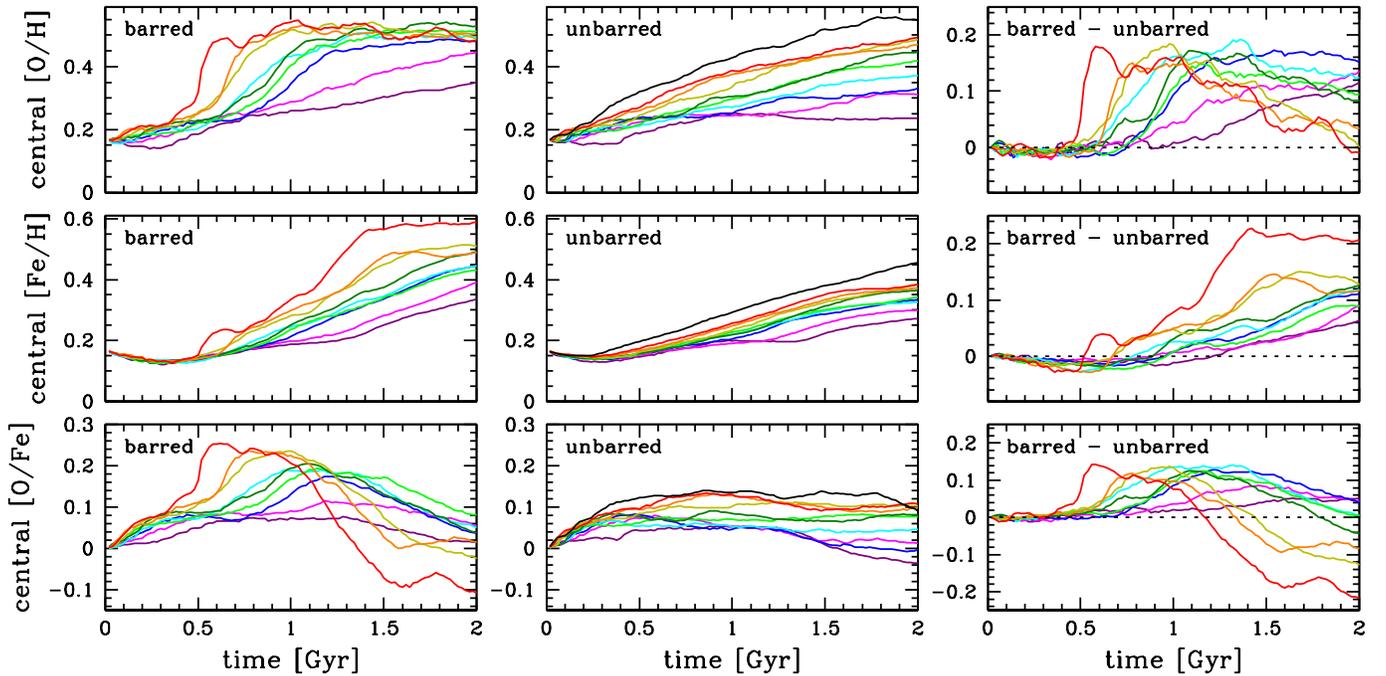}
\caption{Evolution of central values of
\OH\ (top panels), \FeH\ (middle panels), and \OFe\ (bottom panels)
in the gas phase, for galaxies O--H.
Left panels: barred galaxies; central panels: unbarred galaxies;
right panels: difference between barred and unbarred galaxies.
The colours identify the various runs
(see last column of Table~\ref{initial}).
Galaxy I unbarred is included in the central panels (black lines).}
\label{centralXH}
\end{figure*}

Fig.~\ref{centralXH} shows the evolution of
\OH\ (top panels), \FeH\ (middle panels), and
\OFe\ (bottom panels) in the central region in galaxies O--I. 
\OH\ shows a very different trend for high-mass and low-mass 
barred galaxies.
The oxygen abundance
in high-mass barred galaxies evolves fairly rapidly
with \OH\ increasing by $0.3\dex$ between 0.5 and $0.7\Gyr$. It then remains
quite stable for the rest of the simulations. In contrast, low-mass 
galaxies show a slow and steady increase in \OH, 
and intermediate-mass galaxies show a moderate increase,
eventually reaching a plateau at late times.
The oxygen
abundance increases only during the important periods of star formation.
As expected from their SFRs, unbarred galaxies have a gentle 
increase of their central \OH.
In the right
panel, we plot the difference $\Delta\OH$ between barred
and unbarred galaxies.
Low- and intermediate-mass unbarred galaxies have a notably 
lower final \OH\ than their barred counterpart, but this difference
slowly disappears for higher-mass galaxies.
Barred galaxies have, compared to unbarred galaxies,
a quasi-instantaneous increase in oxygen abundance, which climbs by
at least $0.1\dex$ for all galaxies in less than $500\Myr$.
Unbarred galaxies, whose star formation remains stable throughout
the simulation, have thus a continuous increase of their chemical
abundances, for oxygen as well as iron.
Among barred galaxies, the most massive ones show not only an especially rapid
increase of the abundance, but also a levelling of the values of
\OH\ near the end of the simulation.

The evolution of \FeH\ does not proceed as rapidly as that of \OH.
All galaxies, barred and unbarred, show a dip in
\FeH\ during the first $0.3\Gyr$, while only the
lowest-mass galaxies O show a corresponding dip in \OH.
This decrease is caused by gas entering the central region, coming from
regions that are farther from the centre and therefore poorer in
metals.
After $t=0.5\Gyr$ the
iron abundance in barred galaxies increases rapidly, which translates into
a $\Delta\FeH$ of $0.05\dex$ for galaxies F, G, and H.
In lower-mass barred galaxies,
the iron abundance increases progressively
and overtakes the ones of unbarred galaxies. At $t=1.0\Gyr$, barred
galaxies have gas richer in iron in the central region than
unbarred galaxies. While the \OH\ 
of high-mass barred galaxies is not noticeably higher than high-mass 
unbarred galaxies, \FeH\ of barred galaxies is higher than the one 
from unbarred galaxies at all masses.

The evolution of \OFe\ shows an initial increase caused by Type~II SNe,
producing oxygen, followed by a decrease due to Type~Ia SNe
producing iron. Overall, high-mass barred galaxies end up
with significantly lower central values of \OFe\ than the
corresponding unbarred galaxies (by $0.1-0.2\dex$), while in
lower-mass galaxies the differences are much smaller.
In section~1, we stressed the need to compare barred and unbarred galaxies
that share a common observational property related to their
total mass or evolutionary
stage. In this paper, we will use the central stellar mass $M_*$
(stellar mass in the central region of radius $1\kpc$), following
\citet{ellisonetal11}. As Fig.~\ref{centralXH} shows,
the central \OFe\ ratio would be a bad indicator of evolutionary stage.
It varies non-monotonically for barred galaxies, and hardly varies for
unbarred galaxies.

\subsection{Initial gas fraction}

\begin{figure}
\includegraphics[scale=0.45]{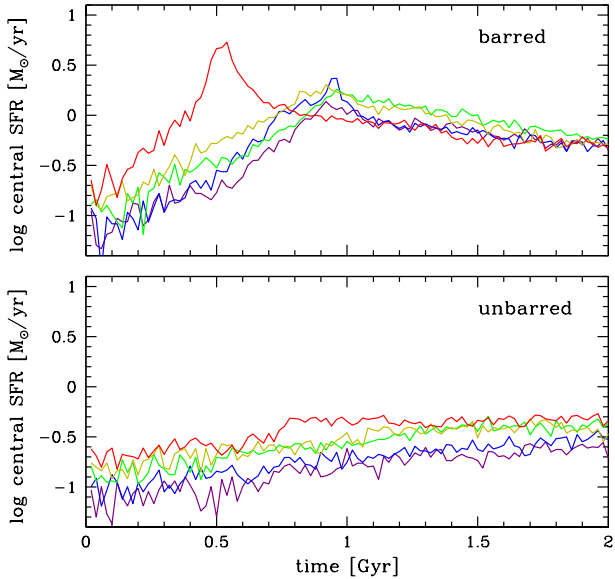}
\caption{Evolution of the central SFR in galaxies $\rm D^{--}$,
$\rm D^-$, D, $\rm D^+$, and $\rm D^{++}$.
Top: barred galaxies; bottom: unbarred galaxies.
The colours identify the various runs
(see last column of Table~\ref{initial}).}
\label{SFRF_D}
\end{figure}

\begin{figure*}
\includegraphics[scale=1.0]{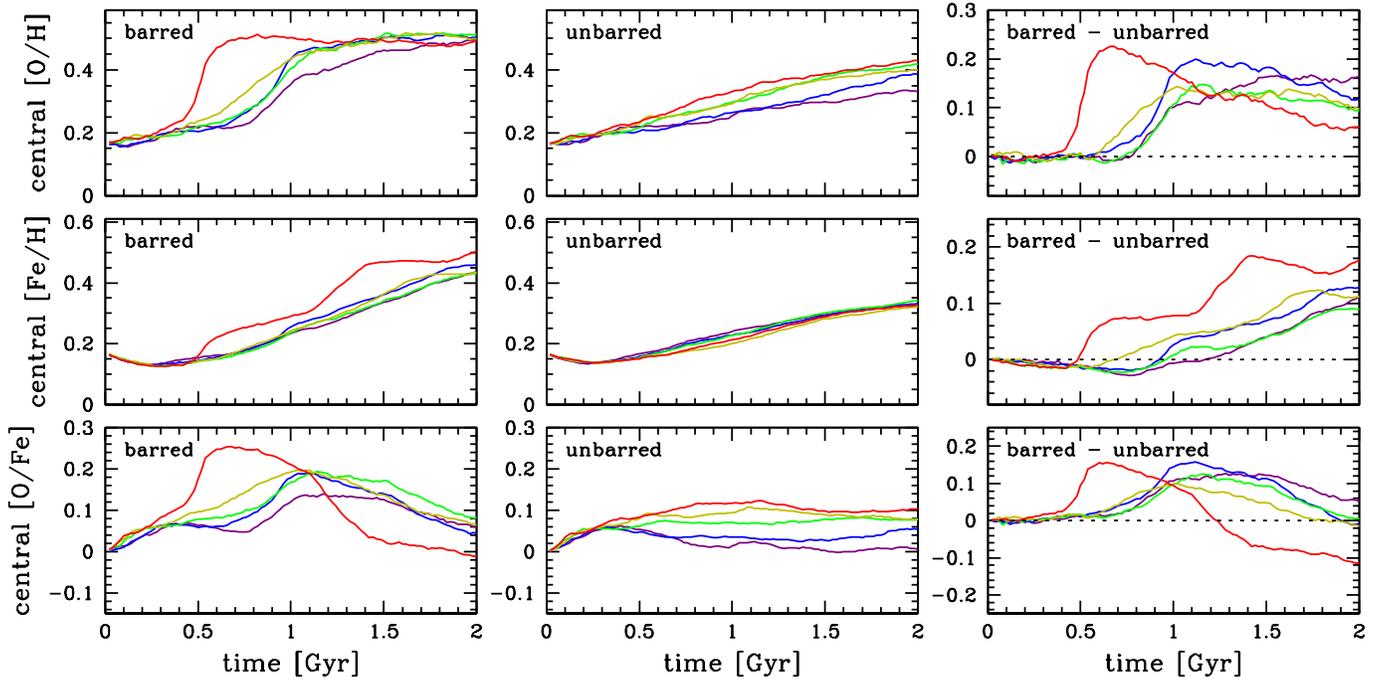}
\caption{Evolution of central values of
\OH\ (top panels), \FeH\ (middle panels), and \OFe\ (bottom panels).  
in the gas phase, for galaxies \DMM, \DM, D, \DP, and \DPP.
Left panels: barred galaxies; central panels: unbarred galaxies;
right panels: difference between barred and unbarred galaxies.
The colours identify the various runs
(see last column of Table~\ref{initial}).}
\label{centralXH_D}
\end{figure*}

In Paper~II, we focused on the effect of stellar mass on the
star formation history
of barred and unbarred galaxies, but we complemented that study
by considering the effect of changing the initial gas fraction.
To do so, we resimulated galaxies D (barred and unbarred),
using two lower initial gas fractions and two higher ones.
These galaxies, named \DMM, \DM, \DP,
and \DPP, are presented in Table~\ref{initial}, with the
initial gas fractions $f_{\rm gas}$ listed in the ninth column.

Figure \ref{SFRF_D} shows the SFR inside the
central region of the various D 
galaxies as a function of time. Barred galaxies from $\rm D^{--}$ to $\rm D^+$
have very similar SFR histories: their SFR increases steadily
between 0 and $1\Gyr$ before decreasing.
The very-gas-rich barred galaxy $\rm D^{++}$ shows a large 
SFR peak around $t=0.5\,\rm Gyr$, similar to the
ones seen in massive barred galaxies F, G, and H 
(Fig.~\ref{SFRF}). As we argued in Paper~II, it is the mass of gas, and
not to total baryonic mass or virial mass which is the primary factor
in determining the star formation history of barred galaxy.
In unbarred galaxies, the SFR varies smoothly, and remains higher in
galaxies with higher initial gas fraction throughout the simulation.

Fig.~\ref{centralXH_D} shows the evolution of
\OH\ (top panels), \FeH\ (middle panels),
and \OFe\ (bottom panels) in the central region
in galaxies \DMM, \DM, D, \DP, and \DPP.
Comparing with Fig.~\ref{centralXH},
we find that the evolution of the abundances in the gas-rich galaxies
\DPP, barred and unbarred, resembles the ones of the massive, gas-normal
galaxies F, G, and H, while the evolution of the other galaxies
\DP--\DMM\ resembles the ones of intermediate-mass galaxies B, C, D, and E.
As for the SFR, it is the gas mass that determines primarily the
evolution of the abundances.

Overall, the gas fraction has little effect
until it gets above a certain value. Then, a small increase in
gas fraction (from 0.290 for galaxy \DP\ to 0.319 for galaxy \DPP)
causes a significant increase both in central
SFR and central metallicities. \citet{silleroetal17} have performed
two simulations of isolated galaxies with different gas fractions
(0.2 and 0.5),
and found that a larger gas fraction leads to
a larger specific SFR and metallicity slope. These authors identify the
formation of gas clumps as the cause for this effect. In our simulations,
the effect is found only in barred galaxies. It would be interesting
to see if the presence of bars favours the formation of clumps, but this is
beyond the scope of this paper and will be considered in future work.

\subsection{The mass-metallicity relation}

\begin{figure*}
\includegraphics[scale=0.8]{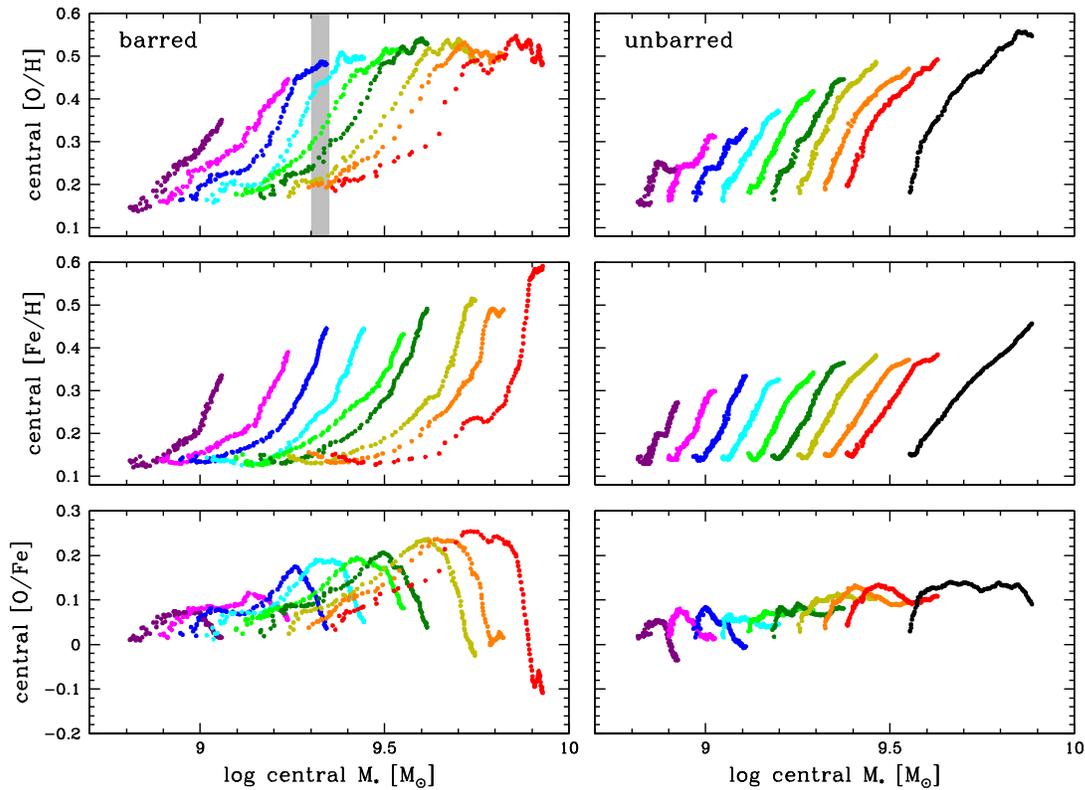}
\caption{Central values of \OH, \FeH, and \OFe\
in the gas phase, for barred galaxies
(left panels) and unbarred galaxies (right panels),
as a function of stellar mass in the central region.
Each dot corresponds to a snapshot of a simulation. Colour indicate
the corresponding galaxies. The gray band in the top left panel
illustrates a mass bin of width $0.05\dex$ centered on a mass
$M_*=2.11\times10^9\msun$.}
\label{M_X2_2Gyr}
\end{figure*}

\subsubsection{Galaxy sample}

In the previous sections, we have examined our results in terms of
``simulation time'' $t$. This is a somewhat arbitrary measure, because the
relation between simulation time and the evolutionary state of an observed
galaxy will depend on our initial conditions, and the evolutionary history of
the observed galaxy. In order to properly compare our results with
observations, we need to parametrise these simulations in terms of
observable variables, as done in, i.e., \citet{scudder_galaxy_2015}. One
limitation of our simulations is our sample size (19 galaxies for runs O--I),
which is much smaller that most samples of observed galaxies. However, each
simulation produces a series a snapshots, which represent the galaxies at
different evolutionary stages. If we consider each snapshot as a separate
galaxy, we then end up with a large sample, covering a wide variety of
physical properties\footnote{with the exception of the dark matter halo
mass $M_{200}$,
which is still limited to the values listed in Table~\ref{initial}.}. We
exclude the first $100$ Myr from our numerical samples to allow the gas to
relax to equilibrium. Note that this is the same procedure we used in Paper~II.

To make meaningful comparison between barred and unbarred galaxies, we
must select galaxies that share a common property. In the previous sections,
we compared galaxies that have reached the same evolutionary time $t$. 
However, this information is not available to observers, who must use a
proxy to identify the evolutionary stages of the galaxies.
In Paper~II, we chose the central stellar mass $M_*$,
a quantity that increases monotonically with time, both in barred and
unbarred galaxies. We use the same approach here. This has the added
advantage of allowing direct comparisons with the observational results
of \citet{ellisonetal11}.

Fig.~\ref{M_X2_2Gyr} shows
the relation between
abundance ratios and central stellar mass.
Each dot represents one snapshot of a simulation, and the colours
identify the runs that provided the snapshots.
The values of \OH\ and central $M_*$ increase with time for all runs.
In barred galaxies, a substantial amount of gas is driven toward
the central region, leading to a rapid increase of both \OH\ and
central $M_*$. Note that some of gas driven toward the centre has been
pre-enriched in oxygen by stars forming outside of the central region
(Paper~I). Overall, barred galaxies reached larger values of
\OH\ and central $M_*$ than unbarred ones (remember that galaxy I unbarred,
shown in black in the right panels, does not have a barred counterpart).
There is a levelling-off
of \OH\ at late time for the most massive galaxies. These galaxies are
at very late stages of their evolution, well passed the peak of star
formation. Stellar evolution produces modest amounts of oxygen at this
point, and the infall of more pristine gas into the central region tends
to compensate that increase, and can even reduce the central value of \OH.

The middle panels of Fig.~\ref{M_X2_2Gyr} shows
the relation between \FeH\ and central stellar mass. Iron is produced by
Type~Ia supernovae, whose explosion is delayed relative to Type~II supernovae.
This explains the small values of \FeH\ in the early evolutionary stages.
Once the late evolutionary stages are reached,
\FeH\ rapidly increases.
In barred
galaxies, the central $M_*$ increases rapidly once the bar forms, and
\OH\ also starts increasing, while there is a significant delay
before \FeH\ starts increasing, explaining the shape of the evolutionary
tracks in the top left and middle left
panels of Fig.~\ref{M_X2_2Gyr}.

The bottom panels of Fig.~\ref{M_X2_2Gyr} show the relation between
\OFe\ and central stellar mass. In the early evolutionary stages,
\OFe\ increases roughly linearly with central $M_*$.
Then, the late production of iron brings the values of \OFe\
down, both in barred and unbarred galaxies. We still find a roughly linear
relation for unbarred galaxies, but not for barred galaxies due to the
large variations in \OFe\ in massive starburst galaxies (lime, orange,
and red dots in bottom left panel of Fig.~\ref{M_X2_2Gyr}).

In our simulations, a mass-metallicity relation in isolated
galaxies naturally appears as a result of secular evolution.
The \OH--$M_*$ relations shown in the top panels of
Fig.~\ref{M_X2_2Gyr} are similar in average values and dispersion
to the observed ones (e.g. \citealt{tremontietal04}, Fig. 6),
except for an overall shift of $\sim0.2\dex$ toward higher values of
\OH. This is a combination of two effects. First, our simulated galaxies
have a significant abundance gradient in the final state, and by considering
only the central $1\kpc$ instead of the whole galaxy, we are shifting the
distribution left (lower $M_*$) and up (higher \OH). Second, we only consider
galaxies that have completed their mass assembly and experience secular
evolution from that point. We are therefore ignoring mass accretion from
the IGM, a process that would tend to reduce the values of \OH.
Until we have a more extended suites of simulations that
includes non-isolated galaxies, we should refrain from making
quantitative comparisons with observations.

\subsubsection{Mass binning}

\begin{figure*}
\includegraphics[scale=0.8]{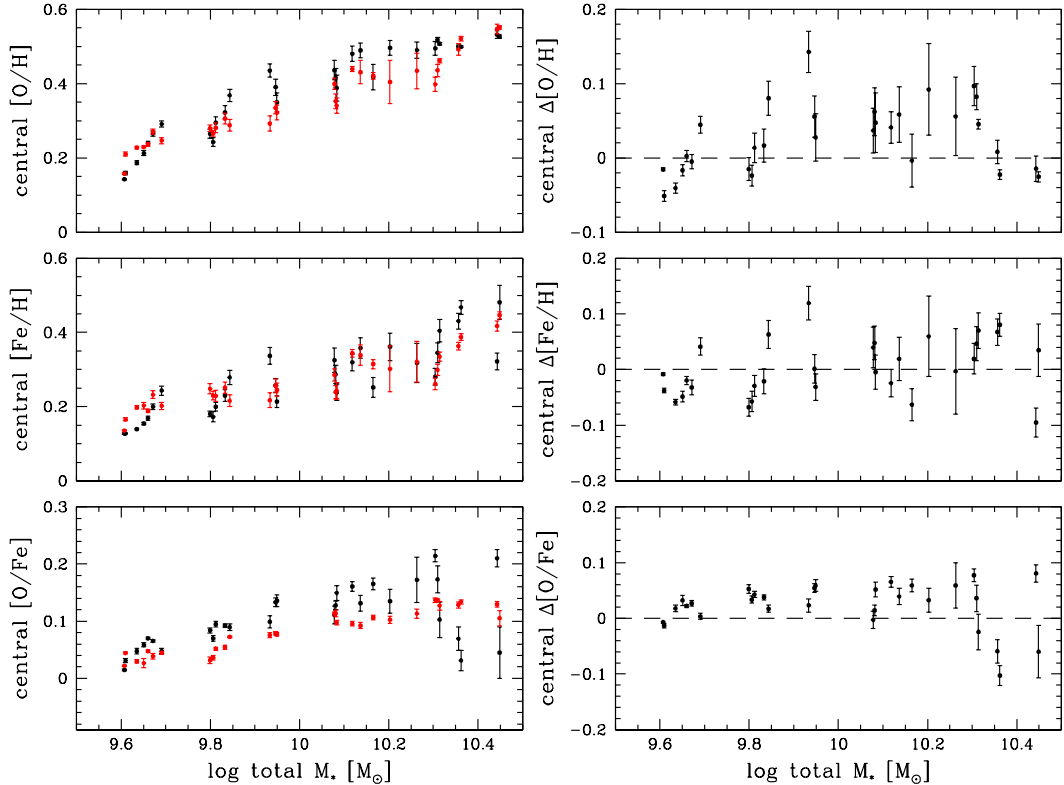}
\caption{
Left panels: central values of \OH, \FeH,
and \OFe\ in the gas phase,
for barred galaxies (black) and unbarred galaxies
(red) having {\it the same
central stellar mass\/} as a function of the {\it total\/} stellar mass of the
barred galaxy.
Right panels : differences $\Delta\OH$, $\Delta\FeH$, and
$\Delta\OFe$ between barred and unbarred galaxies,
as a function of the {\it total\/} stellar mass of the
barred galaxy.
Error bars in the left panels show $1\sigma$ deviations.
Error bars in the right panels were obtained
by adding in quadrature the ones from the left panels.}
\label{M_X_2Gyr}
\end{figure*}

In Paper~II, we noted that galaxies with the same central $M_*$ can have
different global $M_*$. In our simulations the central $M_*$ increases over
time, and so a low-global $M_*$ galaxy at a late stage of evolution can have
the same central $M_*$ as a high-global $M_*$ at an early stage of evolution.
In Paper~II, we examined this effect in terms of the central SFR, but this
affects the central metallicities as well. To illustrate this, we drew in the
top left panel of Fig.~\ref{M_X2_2Gyr} a central stellar mass bin of width
$0.05 \dex$. This mass bin contains high-mass galaxies in their early
evolutionary stages (orange and lime dots), low- mass galaxies in their
final evolutionary stages (blue dots), and everything in between. The
galaxies in this bin cover a range of nearly $2$ Gyr in age, and their
initial stellar masses varies from $6.3\times10^9\msun$ to
$20\times10^9\msun$, yet they have essentially the same central stellar mass.

As in Paper~II, we compare our results with the observations of
\citet{ellisonetal11}, although here we compare metallicities instead of
star formation rates. In \citet{ellisonetal11}, a sample of 294 barred
galaxies was binned by central stellar mass, and the SFRs and metallicities
(in terms of \OH) compared with the expected values of unbarred galaxies
for the same bin. They found that massive ($M_*\geq10^{10}\msun$) barred
galaxies had an average SFR of $0.2 \dex$ higher than that of unbarred
galaxies. In contrast, over all masses they found an enhancement in \OH\
of $0.05\dex$ in barred galaxies compared to unbarred ones. In Paper~II, to
investigate the SFR enhancement in our sample, we recreated the same method
using a $t < 1$ Gyr simulated sample, by calculating a weighted average SFR
in 30 central stellar mass bins for barred and unbarred galaxies. Weights
were selected to represent the relative likelihood of observing a particular
galaxy, determined using the halo mass function of \citet{mpr13}. We found
the low-mass barred galaxies had a small enhancement of $0.2 \dex$ of their
central SFR when compared to unbarred galaxies with equivalent central
stellar mass. This enhancement increases with mass, before reaching a
plateau of $0.4 \dex$ at $\log M_* = 9.9$. Hence, our simulations essentially
reproduced the results of \citet{ellisonetal11}, in particular the location
and amplitude of the transition, except for an overall shift of
$\sim0.2\dex$ at all masses.

We are now applying the same method as in Paper~II,
but calculating the averaged values of \OH, \FeH, and \OFe\ in each mass
bin instead of SFR.
The results are shown in Fig.~\ref{M_X_2Gyr}.
We find an enhancement in metallicity
in barred galaxies compared to unbarred ones between
$\log M_*=9.8$ and $\log M_*=10.3$. For iron (middle right panel),
the results are quite noisy, but if we ignore one data point located
at $\log M_*=10.16$ there is an enhancement of order
$\Delta\FeH\sim0.05\dex$ with a
lot of scatter. For oxygen (top right panel),
there is a clear
enhancement of order $\Delta\OH\sim0.07\dex$ in barred galaxies, except
at the lowest and highest mass bins, where we find no difference.
By contrast, \citet{ellisonetal11} found a uniform enhancement of
$0.05\dex$ in \OH\ at all masses
in the range $\log M_*=9.4-11.2$.
The highest-mass bins contain high-mass galaxies at their late
evolutionary stages. These bins are missing even-higher-mass galaxies
at earlier evolutionary stages, because they were no included
in our series of simulations.
Similarly, the lowest-mass bins contain low-mass galaxies at early
evolutionary stages, but are missing even-lower-mass, more evolved
galaxies.

Finally, the bottom panels of Fig.~\ref{M_X_2Gyr} show the results
for \OFe. There is an enhancement of order $0.4-0.6\dex$ in barred galaxies
at all masses up to $\log M_*=10.3$. At larger masses, the results are
noisy and inconclusive. Finding an enhancement in \OFe\ in an age-limited
sample is not surprising, since the late, iron-producing stages
of some galaxies will always be excluded from the sample.

We note that the large fluctuations in $\Delta\OH$, $\Delta\FeH$, and
$\Delta\OFe$ are caused primarily by large fluctuations in the
values for barred galaxies.
there are sudden jumps in \OH\ and \OFe\ at $\log M_*=10.3$,
There is
a sudden jump in \OFe\ at $\log M_*=10.3$, for barred galaxies. We find
no such sudden jumps for unbarred galaxies.

\subsection{The SFR-metallicity relation}

\begin{figure*}
\includegraphics[scale=0.8]{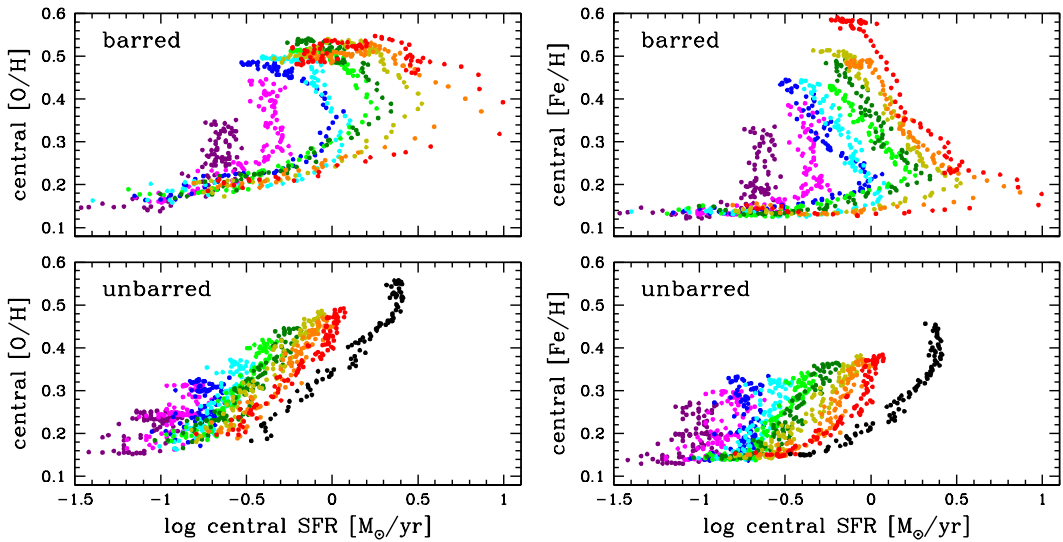}
\caption{Central values of \OH, \FeH, and \OFe\
in the gas phase, for barred galaxies
(left panels) and unbarred galaxies (right panels),
as a function of central SFR.
Each dot corresponds to a snapshot of a simulation. Colour indicate
the corresponding galaxies.}
\label{SFR_X2_2Gyr}
\end{figure*}

According to the basic scenario, bars drive gas toward the centres of
galaxies, where this gas is converted into stars, which later deposit
metals into their surroundings. Hence, we would expect to find a relation
between the central SFR and the central metallicity.
Indeed, the observed mass metallicity relation shows an
additional dependence on SFR
(e.g. \citealt{ellisonetal08b,mannuccietal10,salimetal14}).
We tested the inter-relation between metallicity and SFR rate in
our simulations by using same sample as in the previous section.
Fig.~\ref{SFR_X2_2Gyr} shows the SFR-metallicity
relations. In the early evolutionary stages, there is a general
tendency of \OH\ to increase with the SFR in unbarred galaxies
(bottom left panel). For \FeH\ (bottom right panel), the effect is
weaker. Because of the delay in Type~Ia SNe explosions, each evolutionary
track is initially flat, before \FeH\ finally stars increasing. By
superposing these tracks, we end up with a relation that has a shallow slope
and a large dispersion. At later times, there is
still a tendency of \OH\ and \FeH\ to increase with the SFR, but the
dispersion becomes very large. If we do the same exercise as
in Fig.~\ref{M_X2_2Gyr}, and draw a vertical line in the bottom panels
of Fig.~\ref{SFR_X2_2Gyr}, we would see that a bin at a given SFR contains
galaxies of very different masses, that are at different stages of their
respective evolution: old, low-mass galaxies with high metallicity,
and young, high-mass galaxies with low metallicity.

Within the context of our simulations,
the tendency of \OH\ to increase with SFR makes sense physically.
A higher SFR implies that more stars will form and evolve, and
eventually produce more metals. However, it goes contrary to observations,
which show that at a given $M_*$, \OH\ tends to decrease when the SFR
increases (e.g. \citealt{ellisonetal08b}, Fig.~1;
\citealt{mannuccietal10}, Fig.~1). In both papers, these authors suggest
accretion of low-metallicity intergalactic gas as a possible explanation,
a process which is not considered in our simulations.

The SFR-metallicity relation gets even messier
with barred galaxies (top panels of Fig.~\ref{SFR_X2_2Gyr}).
The evolutionary tracks turn to the left
(lower SFR) because the evolution of the SFR is non-monotonic (see also
Fig.~19 of Paper~I). The metallicity dispersion becomes so large that
there is essentially no SFR-metallicity relation to speak of. Because
of the non-monotonic behaviour of the SFR, which increases before the
starburst and decreases after the starburst, a bin at a given SFR will not
only contain galaxies of different masses at different evolutionary stages,
but also {\it equal-mass galaxies\/} at different evolutionary stages.

The difficulty in relating the central SFR to the central metallicity was
highlighted in Paper~I,
where we showed that a large fraction of metals present in the centre
of barred galaxies did not form in situ, but formed along the entire
length of the bar, to be later transported to the centre by gas flows.
But there is also a conceptual, and more fundamental problem.
The stellar mass and gaseous metallicity are integrated quantities; their
values are determined by the history of star formation and chemical enrichment,
integrated over the life of the galaxies. More specifically, the stellar
mass is the integral of the SFR, while the metallicity is the integral of
the chemical yield. By contrast, the central SFR is an
instantaneous quantity, measured only at the present epoch, and that can differ
wildly from its value at earlier epochs.
Integrated quantities like the stellar mass or central metallicity
tend to vary monotonically with time, while instantaneous quantities like the
central SFR might vary non-monotonically. Consequently, the relationship
between an integrated quantity and an instantaneous one might not
be one-to-one.

\subsection{The effect of AGN feedback}

In this section, we consider the effect of AGN feedback on the evolution
of the metallicity. Runs \AB--\JB\ were presented in Paper~III.
The galaxies have a halo mass $M_{200}=2.306\times10^{12}\msun$,
and most of them have an initial stellar mass $M_*=5.80\times10^{10}\msun$,
consistent with the scaling relations of \citet{belletal03} and
\citet{coxetal06}, the other galaxies
being gas-poor or very-gas-poor.
These masses were chosen for a specific reason:
the shape of the $M_h/M_*$ relation shows that SNe feedback
dominates at stellar masses $M_*<3\times10^{10}\msun$,
whereas most of the gas is blown out of the galaxy by AGN feedback
at much larger masses \citep{bcw10,mosteretal10}.
For this reason, we selected an initial stellar
mass which is near the bottom of the AGN-dominated regime.

\begin{figure}
\includegraphics[scale=0.45]{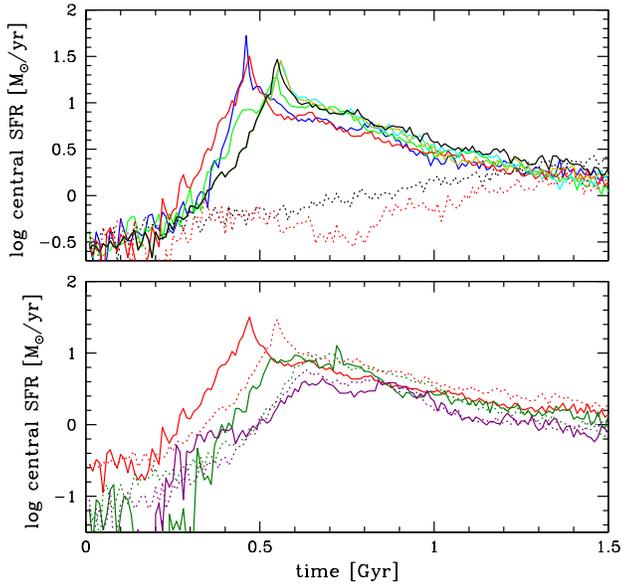}
\caption{Evolution of the central SFR in galaxies \AB--\JB.
Top panel: gas-normal barred galaxies \AB--\FB\ (solid lines),
and gas-normal unbarred galaxies \AB\ and \DB\ (dotted lines).
The colours identify the various runs
(see last column of Table~\ref{initial}).
Note that at $t<0.5\Gyr$, the
black, cyan, and
lime solid lines are superposed, and only the black line is visible.
Bottom panel:
gas-normal barred galaxies \AB\ and \DB\ (red lines),
gas-poor barred galaxies \GB\ and \HB\ (dark green lines)
and very-gas-poor galaxies \IB\ and \JB\ (purple lines);
Solid lines: AGN feedback; dotted lines: no AGN feedback.
}
\label{SFRF_AGN}
\end{figure}

The top panel of Fig.~\ref{SFRF_AGN} 
shows the SFR vs. time inside the central $1\,\rm kpc$ region, for
galaxies \AB--\FB, with solid and dotted lines showing barred and
unbarred galaxies, respectively. 
For the runs without feedback (black lines), the results are consistent
with the ones shown in Fig.~\ref{SFRF}. In barred galaxies, the starburst peaks
earlier in galaxies with feedback (\BB, \CB, \DB) than in galaxies with no
feedback or delayed feedback (\AB, \EB, \FB), even though all galaxies
have the same mass. The presence of feedback favours the growth of a bar
instability, causing the bar to form earlier, which leads to an
earlier starburst
(Paper~III). In Runs \EB\ and \FB, the AGN is turned on at
$t=500\,\rm Myr$, at a time when the bar is formed and
star formation in the central regions is well under way, and the effect of
AGN feedback is then negligible. After $t=800\Myr$, the
central SFR decreases at the same rate in all barred galaxies.
The dotted lines in the top panel of Fig.~\ref{SFRF_AGN}
show the SFR for unbarred galaxies. The central SFR increases slowly,
and the effect of feedback is small except for a brief period
around  $t=800\,\rm Myr$.

The bottom panel of Fig.~\ref{SFRF_AGN} shows the evolution of the SFR
for gas-normal barred galaxies \AB\ and \DB\ (red lines),
gas-poor barred galaxies \GB\ and \HB\ (dark green lines), and
very-gas-poor barred galaxies
\IB\ and \JB\ (purple lines), with solid and dotted lines showing
galaxies with and without AGN feedback, respectively.
The SFR drops significantly with gas fraction, both for galaxies with an AGN
(solid lines) and without (dotted lines).
With a lower
gas fraction, the SFR rises slower, peaks at a lower value, and reaches
this peak later. 
These results are discussed in more details in Paper~III.

\begin{figure*}
\includegraphics[scale=0.8]{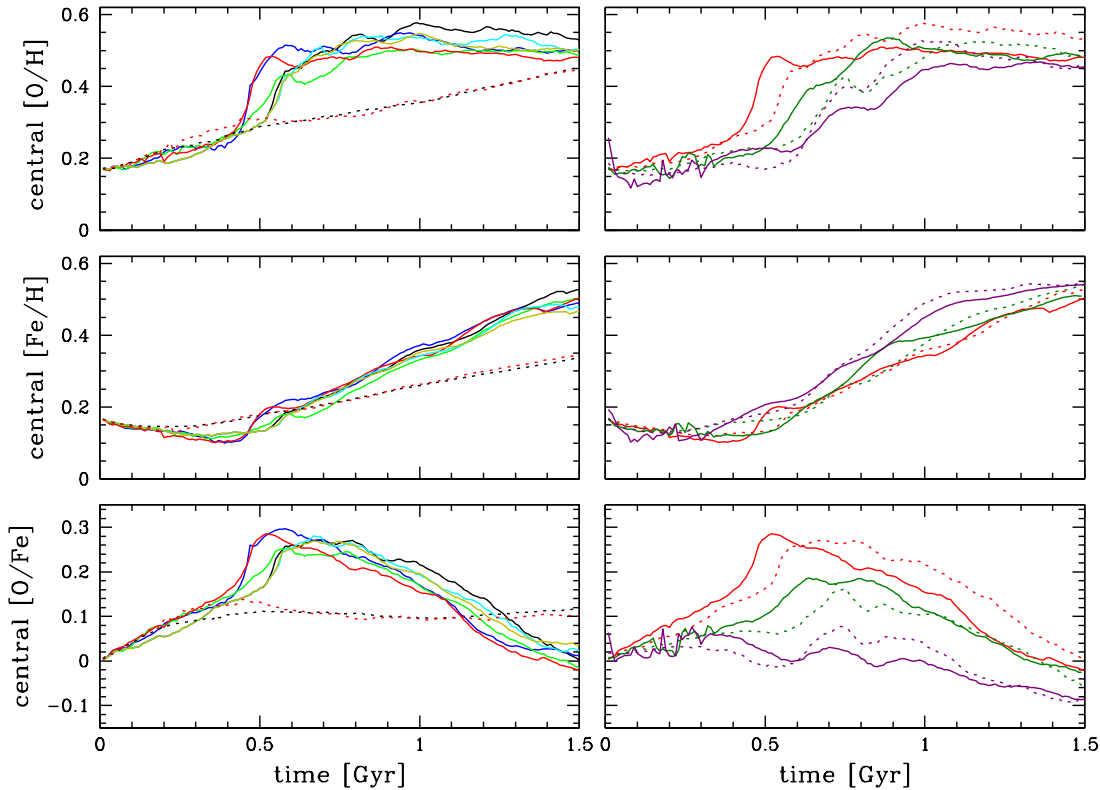}
\caption{Evolution of central values of \OH, \FeH, and \OFe\
in the gas phase, for galaxies \AB--\JB.
Left panels: gas-normal barred galaxies \AB--\FB\ (solid lines),
and gas-normal unbarred galaxies \AB\ and \DB\ (dotted lines).
The colours identify the various runs
(see last column of Table~\ref{initial}).
Right panels: gas-normal barred galaxies \AB\ and \DB\ (red lines),
gas-poor barred galaxies \GB\ and \HB\ (dark green lines)
and very-gas-poor galaxies \IB\ and \JB\ (purple lines); Solid
and dotted lines in right panels show galaxies
with and without AGN feedback, respectively.}
\label{XF_AGN}
\end{figure*}

Fig.~\ref{XF_AGN} shows the evolution of \OH, \FeH, and \OFe. The left panels
show the results for gas-normal galaxies \AB--\FB. All barred
galaxies experience
a starburst, which leads to a sudden increase of $\sim0.25\dex$
in oxygen abundance due
to enrichment by Type~II SNe. Once the starburst is passed, the SFR
rapidly decreases, and the \OH\ ratio levels off
at $\sim0.50\dex$. At late times, \OH\ is $0.05\dex$
larger in the absence of feedback (black line), because the
SFR is larger. Because of the time delay for Type~Ia SNe, the iron
abundance increase slowly, and never reaches a plateau. At 
early times, the iron abundance actually decreases because the inflow
of low-metallicity gas along the bar dilutes the central abundance.
For oxygen, this effect is not sufficient to compensate for the
enrichment caused by stars already present in the initial conditions.
As a result, \OFe\ first rapidly increases, reaching $0.3\dex$ soon
after the peak of the starburst, and then drops progressively,
reaching zero at $t=1.5\,\Gyr$.
For unbarred galaxies, star formation increases very slowly
(Fig.~\ref{SFRF_AGN}), resulting in a slow and continuous enrichment in
oxygen and iron. \OH\ and \FeH\ increase at the same rate, except at
early times when stars present in the initial condition cause an increase in
oxygen relative to iron. \OFe\ increases slowly, and then levels off at
$0.1\dex$. AGN feedback makes hardly any difference in these unbarred galaxies,
because the rate of AGN accretion is quite low, due to a lack of mechanism
for driving gas toward the central regions (Paper~III).

The right panels of Fig.~\ref{XF_AGN} shows the evolution of \OH, \FeH,
and \OFe\ for gas-normal barred galaxies \AB\ and \DB, gas-poor
barred galaxies
\GB\ and \HB, and very-gas-poor barred galaxies \IB\ and \JB. The
SFR is reduced and the starburst takes place later as the gas fraction
goes down, leading to a slower and less sudden increase in \OH.
For gas-normal and gas-poor galaxies (red and dark green lines), the
starburst takes place later when feedback is present
(Fig.~\ref{SFRF_AGN}), delaying the increase in \OH\ by $0.1\,\Gyr$. For
very-gas-poor
galaxies (purple lines), the effect of feedback on the SFR is small, so we
find a different behaviour: \OH\ increases faster in presence of feedback.
\FeH\ increases slowly and steadily, and the values are actually larger
for lower gas fraction. In this case, the dilution caused by infall of
low-metallicity gas dominates over iron enrichment by Type~Ia SNe.
\OFe\ decreases significantly with gas fraction. While \OFe\ reaches
$0.3\dex$ at its peak for gas-normal galaxies, it never exceeds $0.06\dex$
in very-gas-poor galaxies.

\begin{figure*}
\includegraphics[scale=0.8]{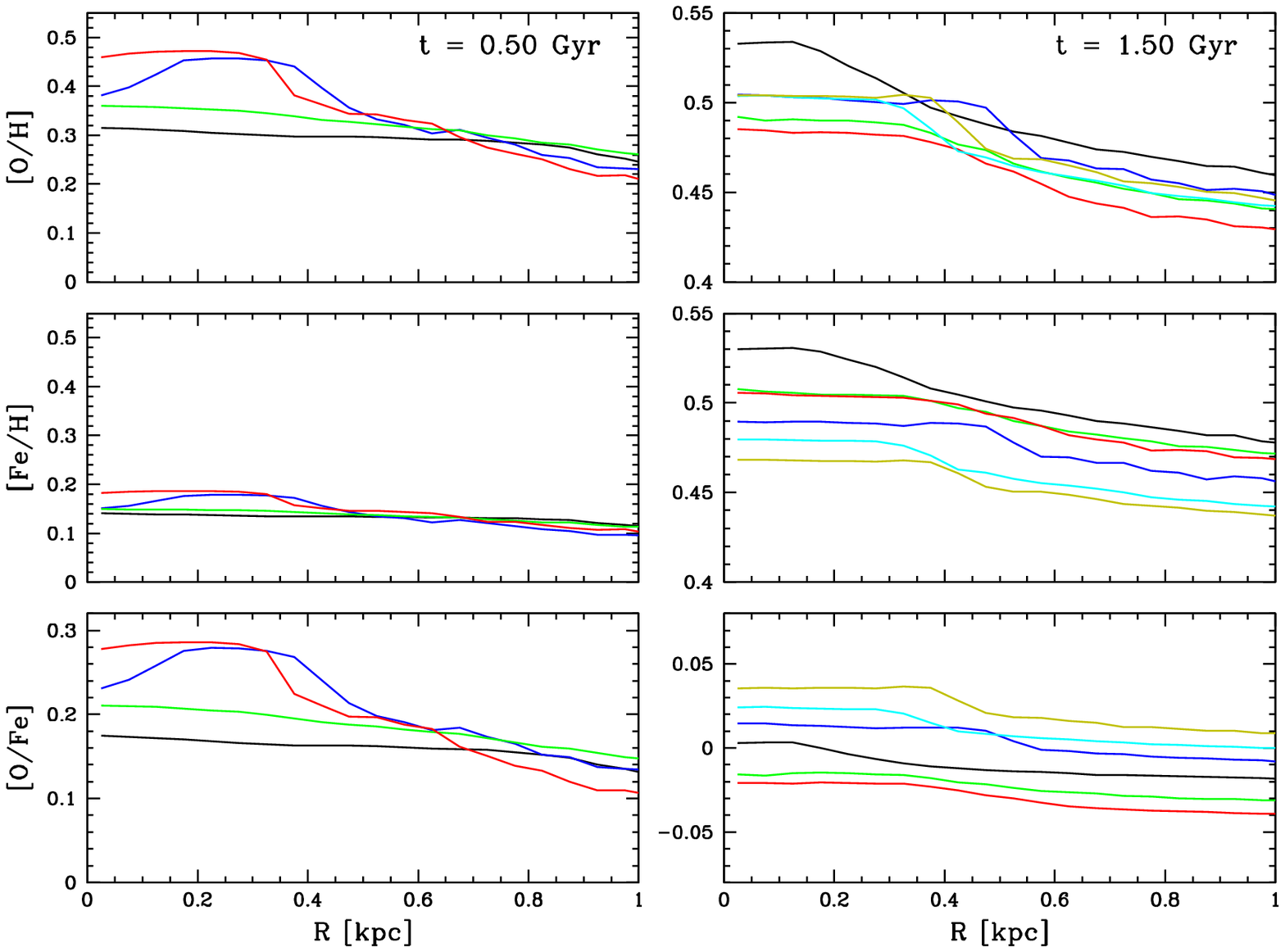}
\caption{Abundance profiles in the gas phase,
in the central region
of galaxies \AB--\FB.
Left panels: profiles at $t=0.5\Gyr$; right panels:
profiles at $t=1.5\Gyr$. The colours identify the various runs
(see last column of Table~\ref{initial}).
In the left panels, black, cyan, and lime lines
(galaxies \AB, \EB, and \FB) are superposed.}
\label{profiles6}
\end{figure*}

\begin{table}
\centering
\begin{minipage}{80mm}
\caption{Mean slope of abundance profiles, for gas-normal galaxies \AB--\FB.}
\begin{tabular}{@{}llccc@{}}
\hline
Time &
Galaxy & $\rm|S(O/H)|$ & $\rm|(Fe/H)|$ & $\rm|S(O/Fe)|$ \\
\hline
$0.5\,\Gyr$ & \AB, \EB, \FB\ &  0.064 & 0.024 & 0.040 \\
            & \BB\           &  0.150 & 0.054 & 0.095 \\
            & \CB\           &  0.098 & 0.036 & 0.061 \\
            & \DB\           &  0.242 & 0.081 & 0.168 \\
\hline
$1.5\,\Gyr$ & \AB\ &  0.072 & 0.051 & 0.021 \\
            & \BB\ &  0.054 & 0.022 & 0.022 \\
            & \CB\ &  0.051 & 0.036 & 0.015 \\
            & \DB\ &  0.055 & 0.036 & 0.012 \\
            & \EB\ &  0.061 & 0.037 & 0.024 \\
            & \FB\ &  0.057 & 0.030 & 0.027 \\
\hline
\end{tabular}
\label{profiles}
\end{minipage}
\medskip
\begin{minipage}{80mm}
Note: Entries show absolute values of slopes, in $\dex\,\kpc^{-1}$.
All slopes are negative.
\end{minipage}
\end{table}

In Paper~III, we found that the effect of AGN feedback in our
simulated barred galaxies did not extend beyond the central region,
for the particular galaxy mass we chose and our assumed AGN model.
The SFR is reduced in the
vicinity of the central black hole, and enhanced at larger radius, up
to $0.4\kpc$. Essentially, feedback pushes the central gas outward, where
it collides supersonicly with infalling gas, forming a dense ring
where star formation is enhanced. Thus, if any chemical
signature of AGN feedback is to be found, it would be
in the radial abundance profiles in the central region. The left panels of
Fig.~\ref{profiles6} show the radial \OH, \FeH, and \OFe\ profiles
in the central region of barred galaxies \AB--\FB\ at $t=0.5\Gyr$,
when the SFR is near its peak. In presence of AGN feedback, the
bar forms earlier, and this modifies the \OH\ profile both at
small and large radius. A bar that forms earlier results in an earlier
starburst, and therefore an earlier oxygen enrichment
of $\sim0.15\dex$, but this effect
is limited to the inner $0.5\kpc$ where most of the star formation
takes place (Paper~III). At larger radius, where the SFR is much lower,
the dominant process is the infall of low-metallicity gas into the central
region, resulting in a slight decrease of $\sim0.04\dex$
in \OH. This process takes place earlier
when the bar forms earlier.

For \FeH, the effect of feedback is much weaker,
and the profiles are much flatter. In Table~\ref{profiles}, we list the
mean slope of the abundance profiles in the central region, in
$\dex\,\kpc^{-1}$. The slopes are 3 times shallower for \FeH\ than
for \OH.
The energy deposited into the central region by the AGN does not couple
with the stellar component. Once gas is converted into stars,
these stars can move through the central region unimpeded before they
explode in Type~Ia SNe, resulting in a fairly uniform distribution
of iron. This results in a \OFe\ profile
that shows an enhancement of $\sim0.05\dex$ in the central
$0.5\kpc$ when feedback is present. The right panels of
Fig.~\ref{profiles6} show the central abundance profiles at $t=1.5\Gyr$.
Two things happen between $0.5\Gyr$ and $1.5\Gyr$. First, the Type~Ia SNe
exploded, allowing the central iron abundance to catch up with the
oxygen abundance, and second, gas motion inside the central
region mixed the elements, leading to much flatter abundance profiles
with slopes shallower than 0.01 (in absolute value).
The differences between galaxies with and without feedback are too
small to be significant.

Overall, for the galaxies considered in this study,
the effect of AGN feedback on the abundance ratios is small and transient.
Differences between the galaxies are important during the
peak of the star formation, and result from the fact that this
peak occurs at different times in the different simulations.
By the end of the simulations, the differences between AGN- and
non-AGN-hosting galaxies, or between AGN-hosting galaxies with different
feedback prescriptions, are insignificant. The effect of AGN feedback on
chemical abundances in indirect, or ``one-off.'' The energy released by the
AGN directly affects the kinematics of the gaseous component,
but all chemical elements are equally affected by this process.
AGN feedback can only affect abundance ratios indirectly,
by (i) modifying the SFR, and the subsequent chemical enrichment history
of the galaxy, or (ii) causing a mixing of regions with different
abundance ratios, differences that originated from processes other
than AGN feedback. One limitation of our study is the fact that,
for galaxies hosting AGNs, we
considered only one particular galaxy mass. This mass was chosen to be
at the lower end of the AGN-dominated regime.
It would be interesting to consider higher-masses galaxies with stronger
AGN activity and feedback.
We defer such study to further work.

\section{SUMMARY AND CONCLUSIONS}

We have conducted a numerical study of the history of chemical enrichment
in barred and unbarred spiral galaxies, focussing on the dependence on
total stellar mass, gas fraction, and AGN activity. The simulations
consist of two separate suites that were presented in previous papers.
The first set consists of 13 barred and 14 unbarred galaxies with
stellar masses in the range $M_*=4\times10^9\msun$ to $50\times10^9\msun$,
and various initial gas fractions. The second set consists of 10 barred
and 2 unbarred galaxies of stellar mass $M_*=58\times10^9\msun$,
with various prescriptions for AGN feedback.
Our simulated barred galaxies experience a sudden increase in central
SFR, followed by a slow decrease caused by gas exhaustion. 
This approach ignores the possibility of replenishing the
supply of gas by accretion from the intergalactic medium,
or mergers with other galaxies. As we explained in Papers~I, II, and~III,
our simulations are most relevant to galaxies located at
sufficiently low-redshift that most of their mass assembly is completed
(e.g. \citealt{lcs12}).
This raises an important question: was the observed
mass-metallicity relation established during the period of mass assembly,
or during the period of secular evolution, or both? By starting the simulations
after the period of mass assembly and using initial conditions with a
fixed metallicity gradient, we are studying the contribution of secular
evolution on the evolution of the metallicity. For this reason,
we reported the changes in abundances during the secular evolution
phase, rather than the actual final values of these abundances.
Our results are the following:

\begin{enumerate}

\item The mass of metals in the central region generally follows a similar
evolution to the total gas mass in the centre. Bars drive gas and
entrained metals toward the centre. The effect is strongest in
massive galaxies. However, these galaxies experience a starburst,
which consumes most of the central gas and metals.
As a result, the most massive barred galaxies eventually end up with a
lower central metal mass and a lower central gas mass.
In unbarred galaxies, the central metal and total gas masses steadily increase,
except for the most massive unbarred galaxy (I), where the high SFR
eventually causes a drop in central gas mass and metal mass.

\item The starburst in massive barred galaxies results in a rapid
increase in central values of \OH.
Once the starburst is passed, the SFR drops and \OH\
levels off. Less massive barred galaxies experience a less sudden
increase in \OH. Eventually \OH\ levels off at the same value in all
barred galaxies except the least massive ones (O, A). In unbarred galaxies,
\OH\ increases slowly and steadily. The central values of \FeH\ in barred
and unbarred galaxies initially decreases because of dilution by infalling
gas. It then increases steadily, with a faster increase in barred galaxies.
The central values of \OFe\ first increase because of oxygen
production by Type~II SNe, then decrease because of iron production
by Type~Ia SNe, both in barred and unbarred galaxies, though the
variations in unbarred galaxies are small.

\item Changing the initial gas fraction has little effect on the evolution
of the central abundance ratios, except for the gas-rich galaxy
\DPP\ barred, which is the only D-galaxy that experiences a starburst.
This galaxy has an initial gas mass comparable to barred galaxies F and G,
which also experienced a starburst. In Paper~II, we concluded that
the initial gas mass was the key factor in determining the star formation
history of the galaxies, and the same is true for the chemical evolution.

\item Using all snapshots provided by our simulations,
we build a mock sample of barred and unbarred galaxies.
With this sample,
we qualitatively reproduced the results of \citet{ellisonetal11},
an enhancement in \OH\ of order $0.7\dex$ in barred galaxies, at
all central stellar masses except at small and large masses, where again
our sample might suffer from incompleteness. We also find, in this
sample, enhancements in \FeH\ of order $0.5\dex$ and in \OFe\ of
order $0.4-0.6\dex$ in barred galaxies.

\item The SFR-metallicity relation shows that the central metallicity
tends to increase with central SFR in all galaxies, up to a point.
If massive, post-starburst barred galaxies are include in the sample,
the SFR does not vary monotonically, and is no longer a good
proxy for identifying the evolutionary stage. In this case,
the dispersion in the SFR-metallicity relation is so large that
this relation is probably useless.
  
\item For the galaxy masses and the AGN feedback model
we considered, the effect of
AGN feedback on the central metallicity is small and transient.
Feedback essentially changes the timing: the bar forms earlier,
leading to an earlier starburst, accompanied by an earlier
oxygen enrichment. Once the galaxies are in their post-starburst
phases, there subsequent evolution are essentially similar. The
final values of \OH\ and \FeH\ in barred galaxies are about
$0.05\dex$ lower when feedback is included. AGN feedback
has no effect of the metallicity of unbarred galaxies, because
in the absence of a bar driving gas toward the centre,
the AGN luminosity is too low to have any effect.

\end{enumerate}

Our results emphases the main difficulty in assessing the effect of
bars on galaxy evolution: deciding which barred and which unbarred galaxies
should be compared to one another. Comparing galaxies with the same
stellar mass $M_*$ is a reasonable (and common) choice, but is not without
problems. Since $M_*$ increases with time in all galaxies, very different
galaxies can have the same value of $M_*$ because they are being observed
at different evolutionary stages. We considered other possibilities,
which turned out to be worse. The SFR and the \OFe\ ratio are
very bad indicators of evolutionary stages, because they tend vary
non-monotonically.
Note that \citet{montuorietal10} and \citet{perezetal11}
reached a different conclusion. However, the simulations of
\citet{perezetal11} did not extend into the regime when iron
enrichment by Type~Ia SNe becomes
important, while the simulations of \citet{montuorietal10} did, but
their analysis focused on the earlier stages evolution, during which
oxygen enrichment by Type~II SNe dominates. Overall,
we reach the same conclusion as in
Papers~I and~II, that it takes at least two observables to estimate
unambiguously at what evolutionary stage a particular galaxy is seen.

Our study shows that the history of metallicity enrichment in barred galaxies
is more complex than the basic scenario suggests.
Our goal was to understand the interplay between
the various physical processes responsible for chemical enrichment
in disc galaxies. Based on our results and the ones presented in Paper~I--III,
we can identify three specific processes that play a key role:

\begin{enumerate}
  
\item In barred galaxies, the presence of a bar drives a substantial amount
of gas toward the central region, while the effect on the stellar
component is weaker (Paper~I). There is exchange of gas and stars between the
central region and its surrounding. As a result, there is no direct connection
between local star formation and local metallicity. Stars, once they form,
can migrate inside the galaxy before exploding, and gas, once enriched, can
flow inside the galaxy from one region to another
(\citealt{dimatteoetal13}; Paper~I; \citealt{kpa13}).
This weakens the significance of any mass-metallicity relation
based on masses and metallicities that are estimated using
central values.

\item There is a time delay between the explosions of Type~II and
Type~Ia SNe, leading to different chemical enrichment histories for iron
and for $\alpha$-elements such as oxygen. The effect is particularly
important in starburst galaxies, where most of the stars form during a
brief period of time. In galaxies with long, steady periods of star formation,
this effect is much weaker.

\item In isolated galaxies, the processes of star
formation and AGN growth compete for the same
limited supply of gas. Each process has the potential to suppress the other
by gas exhaustion.

The potentially most important process not included in
these study is accretion from the IGM. This will be the subject of a
follow-up study.

\end{enumerate}

\section*{acknowledgments}

We are very thankful to Daisuke Kawata for providing useful comments.
This research is supported by the Canada Research Chair program and NSERC.
DJW is supported by European Research Commission grant ERC-StG-6771177
DUST-IN-THE-WIND.

%

\label{lastpage}

\end{document}